\documentclass[prl,10pt,floatfix,amsmath,twocolumn,superscriptaddress]{revtex4-1}
\usepackage[ansinew]{inputenc}
\usepackage{graphicx}
\usepackage{epstopdf}
\usepackage{pst-all}
\usepackage{color}
\usepackage{dcolumn}
\usepackage{amsmath}
\usepackage{amsthm}
\usepackage{bm}
\usepackage{layout}
\usepackage{float}
\usepackage{txfonts}
\usepackage{amsfonts}
\usepackage{amssymb}%
\usepackage[ansinew]{inputenc}
\usepackage{graphicx}
\usepackage{amsmath}
\usepackage{amsthm}
\usepackage{bm}
\usepackage{layout}
\usepackage{float}
\usepackage{amsfonts}
\usepackage{amssymb}
\usepackage{array}
\usepackage{multirow,bigdelim}
\usepackage{enumitem}

\usepackage{CJK}

\usepackage{bm}
\usepackage{graphicx}

\usepackage[colorlinks=true,linkcolor=blue,
pagecolor=blue,filecolor=blue,
menucolor=blue,urlcolor=blue,
citecolor=blue,anchorcolor=blue]{hyperref}%

\begin{document}

\title{Topological Weyl-like Phonons and Nodal Line Phonons in Graphene}

\author{Jiangxu Li}
\affiliation{Shenyang National Laboratory for Materials Science,
Institute of Metal Research, Chinese Academy of Sciences, Shenyang
110016, China} \affiliation{School of Materials Science and
Engineering, University of Science and Technology of China, 110016
Shenyang, Liaoning, China}

\author{Lei Wang}
\affiliation{Shenyang National Laboratory for Materials Science,
Institute of Metal Research, Chinese Academy of Sciences, Shenyang
110016, China} \affiliation{School of Materials Science and
Engineering, University of Science and Technology of China, 110016
Shenyang, Liaoning, China}

\author{Jiaxi Liu}
\affiliation{Shenyang National Laboratory for Materials Science,
Institute of Metal Research, Chinese Academy of Sciences, Shenyang
110016, China} \affiliation{School of Materials Science and
Engineering, University of Science and Technology of China, 110016
Shenyang, Liaoning, China}

\author{Ronghan Li}
\affiliation{Shenyang National Laboratory for Materials Science,
Institute of Metal Research, Chinese Academy of Sciences, Shenyang
110016, China}

\author{Zhenyu Zhang}
\affiliation{International Center for Quantum Design of Functional
Materials (ICQD), Hefei National Laboratory for Physical Sciences at
the Microscale, and Synergetic Innovation Center of Quantum
Information and Quantum Physics, University of Science and
Technology of China, Hefei, China}

\author{ Xing-Qiu Chen}
\email{xingqiu.chen@imr.ac.cn} \affiliation{Shenyang National
Laboratory for Materials Science, Institute of Metal Research,
Chinese Academy of Sciences, Shenyang 110016, China}

\date{\today}

\begin{abstract}
By means of first-principles calculations and modeling analysis, we
have predicted that the traditional 2D-graphene hosts the
topological phononic Weyl-like points (PWs) and phononic nodal line
(PNL) in its phonon spectrum. The phonon dispersion of graphene
hosts three type-I PWs (both PW1 and PW2 at the BZ corners \emph{K}
and \emph{K}', and PW3 locating along the $\Gamma$-\emph{K} line),
one type-II PW4 locating along the $\Gamma$-\emph{M} line, and one
PNL surrounding the centered $\Gamma$ point in the $q_{x,y}$ plane.
The calculations further reveal that Berry curvatures are
vanishingly zero throughout the whole BZ, except for the positions
of these four pairs of Weyl-like phonons, at which the non-zero
singular Berry curvatures appear with the Berry phase of $\pi$ or
-$\pi$, confirming its topological non-trivial nature. The
topologically protected non-trivial phononic edge states have been
also evidenced along both the zigzag-edged and armchair-edged
boundaries. These results would pave the ways for further studies of
topological phononic properties of graphene, such as phononic
destructive interference with a suppression of backscattering and
intrinsic phononic quantum Hall-like effects.
\end{abstract}


\maketitle Graphene, consisting of the one-atom-thick carbon in the
two-dimensional hexagonal lattice, distinguishes itself as an ideal
platform for various interesting and unusual properties, such as
large, tunable carrier densities $n$~$\sim$~10$^{11}$-10$^{14}$
cm$^{-2}$, an ultrasmall, tunable Drude mass, and exceptionally long
intrinsic relaxation times, and so on. These are mainly because its
electronic structure can be described at low energies by a massless
Dirac-Fermion model. On basis of graphene, Kane and Mele contributed
a pioneered theoretical discovery to predict its quantum spin Hall
effect with breaking the Dirac cone into a gap by forcing its
spin-orbit coupling interaction \cite{Kane2005}, witnessing a new
phase of quantum matter, lately called topological insulators in
HgTe\cite{Zhang2006,Zhang2007}, with insulating bulk and quantized
and robust edge conductance. In parallel with electrons, still on
basis of the structure of graphene the topological nature of the
phonon Hall effect was theoretically proposed by interplaying
Raman-type spin-phonon interaction~\cite{Zhanglifa2010} and the
infrared topological plasmons was also recently proposed by breaking
time-reversal symmetry under a static magnetic
field~\cite{Jindafei2017}. The other two time-reversal-symmetry
breaking two-dimensional systems were theoretically proposed to show
topological phonon states with robust one-way elastic edge waves
\cite{Pordan2009,Wang2015}, which immune to backscattering.
Interesting, in all these
studies~\cite{Kane2005,Zhanglifa2010,Jindafei2017,Pordan2009,Wang2015}
the time-reversal-symmetry breaking fields by gyroscopic inertial
effects~\cite{Wang2015,Pordan2009}, spin-orbit coupling
effects~\cite{Kane2005}, spin-phonon
interaction~\cite{Zhanglifa2010}, and static magnetic
field~\cite{Jindafei2017} are necessary to induce the topologically
protected one-way electronic or phononic edge states on 2D systems.
But, to date it has not been still clear whether or not the phonon
spectrum of graphene itself is topological.

In Ref.~\cite{Wang2015}, the modeling analysis for four phonon bands
with the occurrence of the complete band gap for a hexagonal
phononic crystal reveals that this gap is topologically trivial,
since the time-reversal symmetry is not broken and the Chern numbers
of the bands are all zero. It was the reason as to why the authors
introduced gyroscopic coupling to their modeling to obtain the
non-trivial topological nature. In similarity to this work, in
Ref.~\cite{yzliu2017} various novel topological effects of phonons,
including topologically protected pseudospin-polarized interface
states and phonon pseudospin Hall effect, have been theoretically
modeled in the Kekul$\acute{e}$ lattice. Returning to graphene, its
unit cell has two carbon atoms allowing six degrees of freedom for
atomic displacements. Even with the equal masses for two carbon
atoms, it is possible to have intrinsically topologically protected
phononic states in graphene because the extra two freedoms --
vibration modes -- along the direction normal to the \emph{xy}
plane, that definitely increases its perturbations. With such a
purpose, we have revisited the issue of the phonon dispersions of
graphene. Interestingly, we have found that the topology is the
intrinsic property of the phonon spectrum for graphene. Our
calculations reveals that, in the 2D hexagonal Brillouin zone (BZ),
Weyl-like phonons not only exist at two inequivalent ${K}$ and
${K}^\prime$ points but also appear on the $\Gamma$-L line and
$\Gamma$-K line. Both of type-I and type-II Weyl-like phonons are
found in the BZ. More interesting, there still exists a phononic
nodal line (PNL) surrounding the centered ${\Gamma}$ point.
Furthermore, we have evidenced the non-trivial edge states along
both zigzag-edged and armchair-edged boundaries, which are indeed
confined to the boundaries in one-way propagation.

Based on the density functional
theory(DFT)~\cite{Hohenberg.P1964,Kohn.W1965} and density functional
perturbation theory(DFPT)~\cite{Baroni.S2001}, we have calculated
the stable lattice constants and the phonon dispersion. Both DFT and
DFPT calculations have been performed by the Vienna \emph{ab initio}
Simulation Package
(VASP)~\cite{G.Kresse1993,G.Kresse1994,G.Kresse1996}. We adopted the
projector augmented wave (PAW)~\cite{P.E1994,G.Kresse1999}
potentials and the generalized gradient approximation (GGA) within
the Perdew-Burke- Ernzerhof (PBE) exchange-correlation
function~\cite{J.P1996}. They treat semi-core valence electrons as
valence electrons. To obtain the stable phonon spectra, a high
accurate optimization of the lattice constants have been performed
by minimizing the interionic forces within 0.0001 ev/${\AA}$. The
cut-off energy for the expansion of the wave function into the plane
waves was 550 eV. The Monkhorst-Pack k-meshe (21$\times$21$\times$1)
is used for the BZ integrations with a resolution of
2$\pi\times$0.01$\AA$. By trying various supercells for the
calculations of phonon dispersions, it has been found that the
7$\times$7 $\times$1 supercell yielded a high accuracy to provide
the most reliable force constants by combining both the VASP and
Phonopy code~\cite{L.Chaput2011}. Of course, we have also computed
the phonon spectrum by including the spin-orbital coupling (SOC)
effect and the SOC doesn't exhibit any influence on phonon. By using
the force constants as hopping parameters, we have built the dynamic
matrices to analyze the topological nature. The boundary phonon
dispersions have been performed by constructing the chain model and
the boundary-edged phonon densities of states have also been
obtained by using the iteration Green's function
method~\cite{M.P1985}.

\begin{figure}
\includegraphics[width=0.47\textwidth]{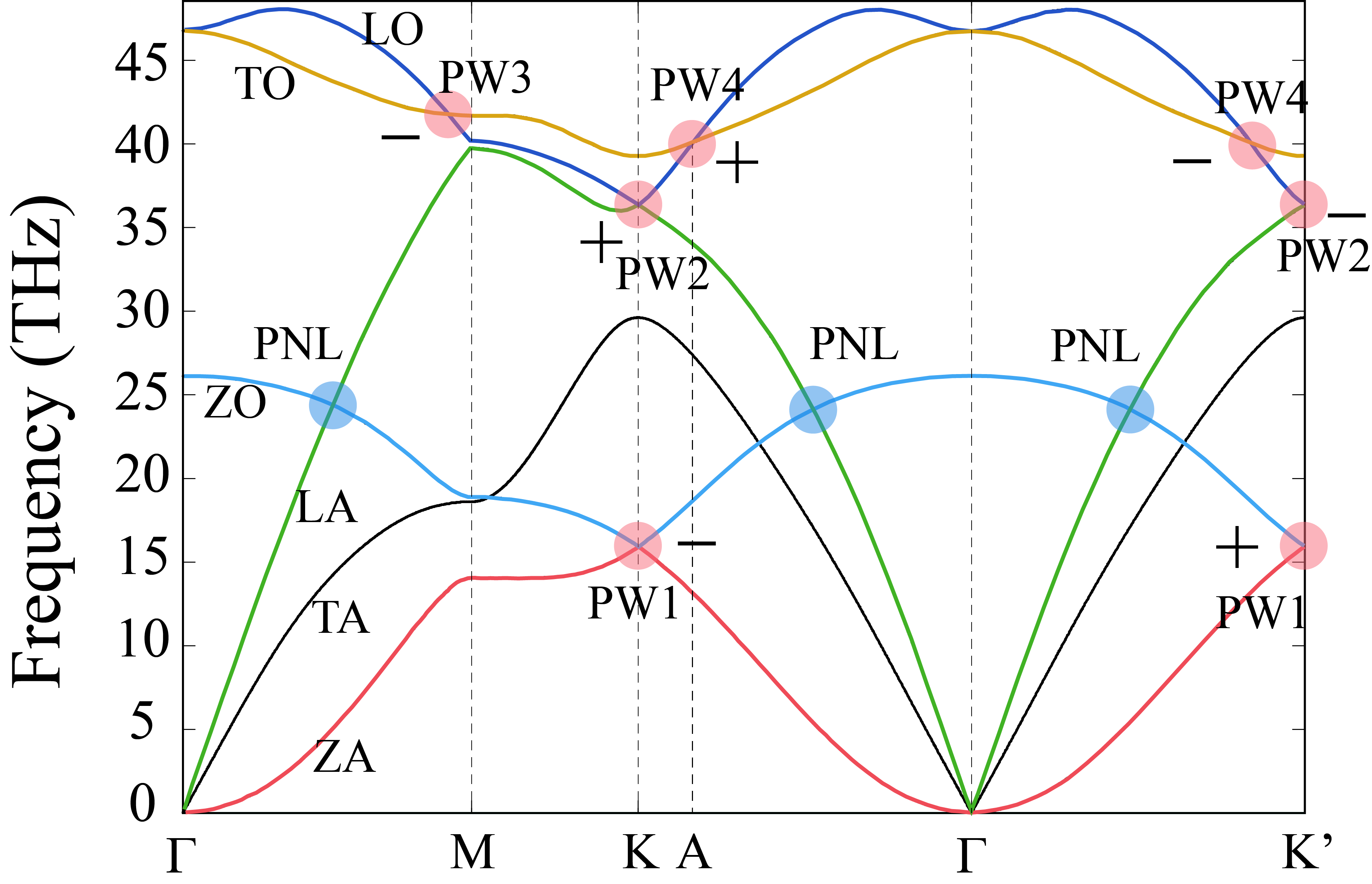}
\caption{The DFT-PBE derived phonon dispersions of graphene. ZA: the
out-of-plane acoustic branch, ZO: the out-of-plane optical branch,
TA: transverse-acoustic branch, TO: transverse-optical branch, LA:
longitudinal-acoustic branch, LO: longitudinal-optical branch. Note
that seven phononic Weyl nodes are marked by the pink transparent
circles, and the three blue circles are three nodal points on the
nodal line formed by the crossing between the lowest out-of-plane
optical (ZO) and the highest longitudinal-acoustic (LA) branches.}
\label{fig:n1}
\end{figure}

\begin{figure*}
\includegraphics[width=0.80\textwidth]{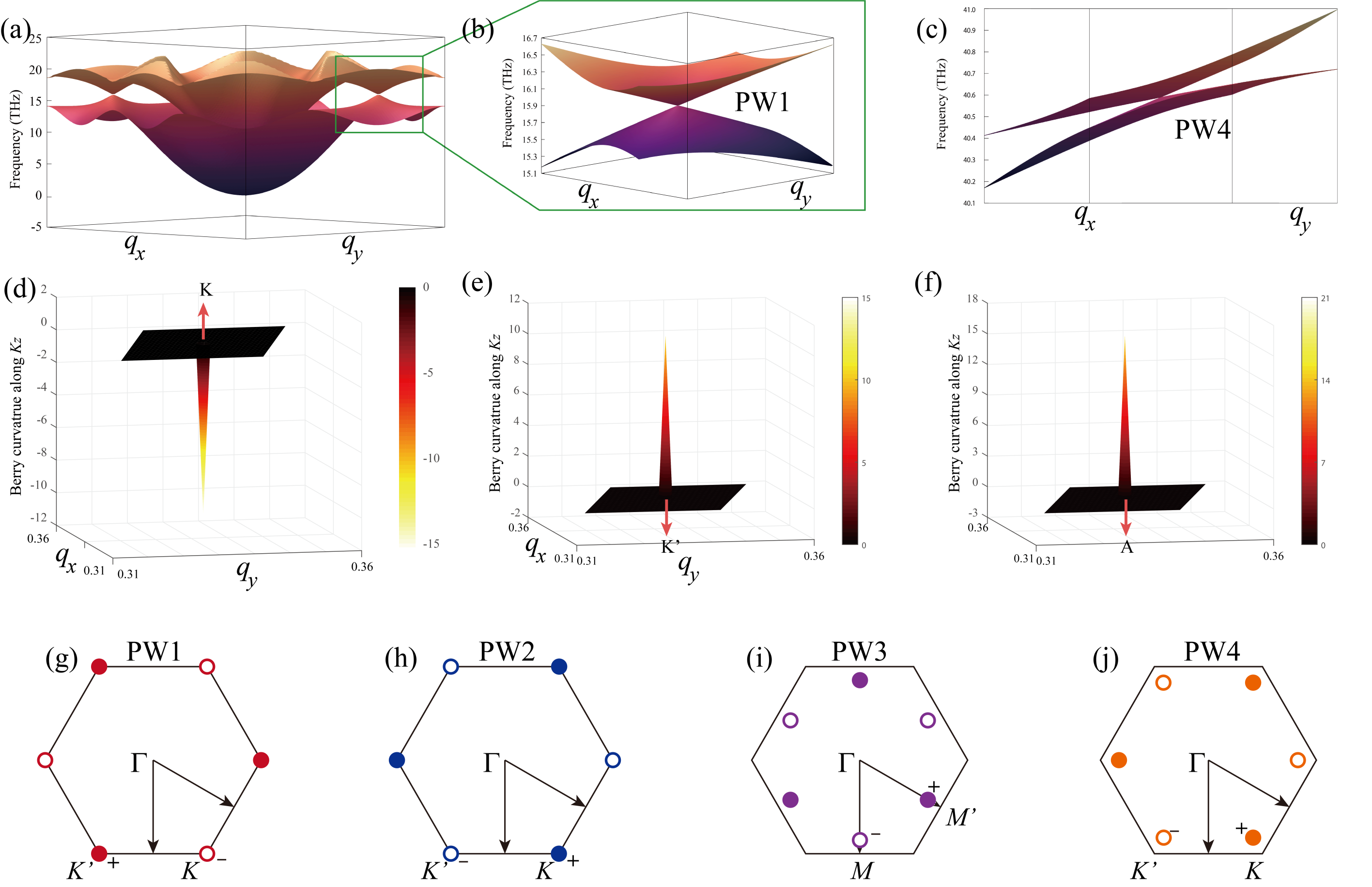}
\caption{Weyl-like phonons at $K$ and $K^\prime$ in graphene. Panel
(a): The 3D visualization of the DFT-PBE derived phonon ZA and ZO
branches of graphene to show the linear crossing, PW1, at the $K$
and $K^\prime$ point in the BZ. Panel (b): The zoom-in 3D
visualization of both ZA and ZO branches surrounding the \emph{K}
point of the BZ. Panel (c): The visualized DFT-PBE derived phonon LO
and TO branches of graphene to show the linear crossing, PW4, along
the $\Gamma$ and $K$ path in the BZ. Panel (d and e): The derived
Berry curvature surrounding the phononic Weyl-like point, PW1, at
\emph{K} (or $K^\prime$) point on the $q_{x,y}$ plane ($q_x$ = 0.31
-- 0.36; $q_y$ = 0.31 -- 0.36) of the BZ. Panel (f): The
distribution of Berry curvature of the Weyl point, PW4, at the
defined A point of the 2D BZ. Panels (g to j): the distribution of
the Weyl-like points in the first BZ of graphene, PW1, PW2, PW3 and
PW4, respectively.} \label{fig:n2}
\end{figure*}

We have recalculated the phonon dispersions of graphene in
Fig.~\ref{fig:n1}, which is in nice agreement with previous
calculations ~\cite{Mounet2005,Bonini2007,Nika2009}. Importantly, we
have observed several linear crossings of the phonon bands in
Fig.~\ref{fig:n1}. At the $K$ point - the BZ concers, there are two
linear crossings of the phonon bands, clearly stemming from the band
crossing between the ZA and ZO branches at 15.92 THz and from
another band crossing between the LA and LO branches at about 36.36
THz. The former is marked with PW1 and the latter with PW2 in
Fig.~\ref{fig:n1}. In addition, we still note that there exist the
two types of band crossings between LO and TO branches along
$\Gamma$-\emph{M} and $\Gamma$-\emph{K} (or $\Gamma$-$K^\prime$)
paths, as defined as PW3 and PW4, respectively. In particular, these
four crossing points (PW1, PW2, PW3 and PW4) are the isolated
points, showing the conical band structure on the $q_{x,y}$ plane.
The PW1 has the lowest frequency among those four crossings. We have
plot the band structure between the ZA and ZO branches on the
$q_{x,y}$ plane of the BZ in Fig.~\ref{fig:n2} (a and b) from which
the conical structure can be clearly evidenced. Interestingly, the
conical shapes of the PW2 and PW3 are very similar to that of the
PW1, although their frequencies are different. However, the conical
shape of the PW4 highly differs, as its conical structure on the
$q_{x,y}$ plane is tilted as shown in Fig.~\ref{fig:n2}(c). To
identify whether these four PWs exhibit topological nature, we have
calculated their Berry curvatures on a 2D $q_{x,y}$ plane. It needs
to be emphasized that for 2D crystals the Berry curvature only has
non-zero value ($\Omega_{xy}$) along $q_z$ direction, whereas
$\Omega_{yz}$ and $\Omega_{zx}$ have to be zero. Here, we have
selected PW1 and PW4 as the examples. Figure~\ref{fig:n2}(d and e)
shows the Berry curvatures of the ZA/ZO crossing (PW1) at $K$ and
$K^\prime$, indicating that their Berry curvatures only have the
extremum exactly at $K$ and $K^{\prime}$ but with the opposite signs
and at all other position the Berry curvature is strictly zero. The
fact reveals that the PW1 at the $K$ and $K^{\prime}$ have opposite
charges. Furthermore, both PW2 and PW3 show the similar feature to
PW1. We have also calculated the Berry curvature of the PW4 on the
2D $q_{x,y}$ plane in Fig.~\ref{fig:n2}(f), evidencing the isolated
maximum value only at the defined \emph{A} point along the
$\Gamma$-$K$ path, whereas the Berry curvature at any other \emph{q}
point is almost zero. Moreover, we have calculated their Berry
phases as
\begin{equation}
\gamma_n = \oint_C \mathbf{A}_n(\mathbf{q}) \cdot d\mathbf{l},
\end{equation}
where $\mathbf{A}_n(\mathbf{q}) = i\langle u_n(\mathbf{q}) |
\nabla_q |u_n(\mathbf{q}) \rangle$ is the Berry connection and
$u_n(\mathbf{q})$ is the Bloch wavefunction of $n$-th band. For this
purpose, we have defined a closed circle on the $q_{x,y}$ plane
centered at the $q$ = $K$ momentum to calculate Berry phase. The
radius of the closed circle going around this \emph{K} point can be
selected to be arbitrary large, as long as it does not also cover
another $K$ or $K^\prime$ point. Interestingly, the Berry phase for
this crossing point PW1 at $K$ is $-\pi$, whereas another crossing
point at K$^\prime$ has an opposite Berry phase of $\pi$. This fact
means that the two crossing points of PW1 at $K$ or $K^\prime$ are
topological non-trivial and also proves that the topological
property of the band crossings at $K$ or $K^\prime$ are opposite in
their chirality. Therefore, these band crossing points, PW1, at $K$
and $K^\prime$ are indeed a pair of phononic Weyl-like nodes with
the opposite chirality. Thus, we have also analyzed the Berry phases
of the other three crossing points (PW2, PW3 and PW4), revealing
that they have the same topological property as PW1 with the value
of $\pi$ or $-\pi$. Therefore, all these four crossing points of
PW1, PW2, PW3, and PW4 are phononic Weyl-like points. It needs to be
emphasized that the general Weyl point is in the 3D conical
shape~\cite{Weyl1929}, as already emphasized in many electronic Weyl
semimetals~\cite{Wan2011,Huang2015,Weng2015,Xu2015a,Xu2015b,Soluyanov2015,
Sun2015,Yu2017,Fang2012,Dubcek2015,Lu2014,Lu2013,Lin2016,Chen2016,Noh2017,
Hailong2018,Xiao2015,Yang2015,Belopolski2016,Ahn2017,XuYoug2015,Rocklin2016,
Yan2016,Ruan2016,Gao2016,Lu2015,Lv2017,Gooth2017,Inoue2016,Weng2016a,Weng2016b,Wang2017}
and several phononic Weyl
materials~\cite{Stenull2016,Zhang2017,hmiao2018,ljx2018,qxie2018,SS2018,xie2019}.
However, given the fact that it is not physically allowed for any 3D
shape for graphene due to its single layer structure and the
phononic band is spinless, it is reasonable to define these
crossings as Weyl-like phonons. Given the fact that PW1, PW2 and PW3
exhibit the normal conical shape whereas the PW4 has tilted conical
shape, PW1, PW2 and PW3 are the so-called type-I Weyl-like points
and PW4 is the typical type-II Weyl-like point. All these Weyl-like
nodes obey three-fold rotation symmetry and thus each type of
Weyl-like points have three pairs in their 2D first BZ, as
illustrated in Fig.~\ref{fig:n2}(g, h, i and j).

\begin{figure}
\includegraphics[width=0.47\textwidth]{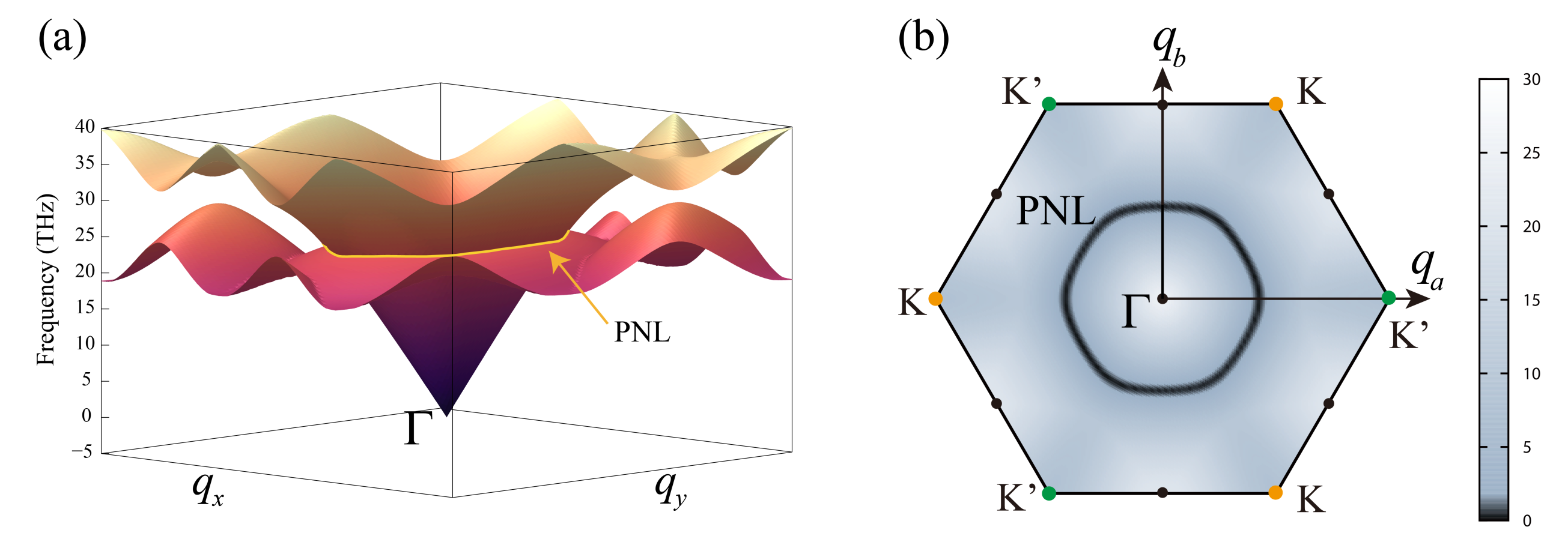}
\caption{Panel (a): The 3D visualization of the DFT-PBE derived
phonon ZO and LA branches of graphene to show the phononic nodal
line surrounding the $\Gamma$ point in the BZ. Panel (b): The gap
between these two ZO/LA branches in the first BZ. The black closed
circle represents the PNL on $q_{x,y}$ plane.} \label{fig:n3}
\end{figure}

We have still noted that at 24 THz the phononic linear band
crossings occur between the LA and ZO branches along the $\Gamma$-K
(or $K^\prime$) and $\Gamma$-$M$ directions (Fig.~\ref{fig:n1}).
Importantly, these three linear crossing points are not isolated
and, instead, they form a closed phononic nodal line (PNL) around
$\Gamma$ point in the BZ. In order to clearly see the shape of the
PNL, we have visualized the phonon bands of both ZO and LA branches
on the $q_{x,y}$ plane in Fig.~\ref{fig:n3}(a). The specified
location of the PNL is shown in the BZ in Fig.~\ref{fig:n3}(b) in
which we have used the gap between the ZO and LA branches to obtain
the PNL distribution, evidencing the occurrence of the closed black
circle centered at the $\Gamma$ point.

For 2D crystal, the topological non-trivial nature can be observed
by the edge states. To elucidate this feature, we have employed the
phononic tight-binding model to construct the supercell of a ribbon
model (Fig.~\ref{fig:n3}(a)) along different directions and the edge
phononic bands are obtained by the truncated chain of graphene.
Through the Green's function iteration method, the 2D and edge
phononic densities of states can be obtained through the imaginary
parts of the Green's function. Currently, we have mainly focused on
the two representative edge states of the zigzag-edged boundary (the 
[100] direction in Fig.~\ref{fig:n4}(a)) and of the armchair-edged
boundary (the [1$\bar{1}$0] direction) as shown in supplementary
materials~\cite{SM}. From the phonon dispersions of the zigzag-edged
boundary in Fig.~\ref{fig:n4}(b), four distinct topologically
protected non-trivial edge phononic states can be observed. In the
first, at about 16 THz of the zigzag-edged phononic states a
straight-line state is formed to connect two projected type-I PW1
points with the opposite charity. To illuminate the edge states, we
have analyzed the Bl\"och modes of a selected point on the edge
states. As shown in Fig~\ref{fig:n4} (b) and (e), the point is
marked by the blue circle on the edge states and its coordinate is
$q$ = (0.45, 0.0)$\frac{2\pi}{a}$. This two-fold degenerate band
originates from the ZA and ZO modes of the pristine phonons of
graphene. From the tight binding model, we have analyzed the
vibration modes of these two phononic bands on the edge states. As
expected, these two edge bands are associated with the out-of-plane
vibration modes only along the $k_z$ direction for the edged carbon
atoms in Fig.~\ref{fig:n4}(a), which are exactly the same as the ZA
and ZO modes correlated with the PW1 Weyl nodes of pristine
graphene. In particular, at $q$ = (0.5, 0.0)$\frac{2\pi}{a}$ each of
the twofold degenerate phonon edge bands is only contributed by the
carbon atoms at the zigzag boundaries. As illustrated in
Fig.~\ref{fig:n4}(a), the sharp spatial Bl\"och modes of those two
phononic bands are strictly confined at the zigzag edges. This fact
means that the two-fold degenerate edged states only propagate along
the edged boundaries. Remarkably, this straight-line phononic edge
state has a nearly flat phononic dispersion upon various $q$
momentum from $\overline{K}$ to $\overline{K^\prime}$, as
illustrated in Fig.~\ref{fig:n4}(e). It reveals that the phononic
edge states originated from the PW1 Weyl-like points have the nearly
zero phonon group velocity (namely, $v$($q$) $\approx$ 0) in the
phonon one-way transport direction along the zigzag-edged boundary
of graphene. This interesting feature of the zigzag-edged narrow
ribbon of graphene would have potential application for the emitting
of the ultra slow light. Of course, we have also analyzed other
points of the phononic edge states, revealing the similar spatial
and local properties of vibrations. In the second, for PW2 at 36.36
THz there is a distinct two-fold edge state connecting two
PW2-projected points with different chirality as shown in
Fig~\ref{fig:n4}(b), which is marked by red and blue lines. At
around 41.50 THz, the other phonon edge states can be also
visualized to connect type-I PW3 Weyl-like points with the opposite
charity, as shown in Fig.~\ref{fig:n4}(c). In comparison with the
edge states induced by PW1 and PW2, the PW3-induced edge states are
not only shorter in the \emph{q} momentum but also exhibit
relatively larger dispersions. Moreover, it needs to be emphasized
that for PW4 no edge states can be clearly observed because the
PW4-induced topologically protected phononic edge states fully
overlap with the edge projection from the phonon dispersions of the
pristine graphene. In the third, in similarity to the nodal lines in
3D crystals which exhibits non-trivial drumhead-like surface states
for BaSn$_2$, Ca$_3$P, TlTaSe$_2$ and TiSi,
ZrSiS~\cite{Huang-PRB-2016,Li2018,Chan-PRB-2016,Bian-PRB-2016,Madhab2016}
and the phononic Weyl nodal line MgB$_2$~\cite{qxie2018}, the
existence of PNL of the pristine graphene at 24.38 THz induces the
straight-line edge states, simultaneously, with an arc-like
going-downwards parabola in its frequencies upon the $q$ momentum,
as shown in Fig.~\ref{fig:n4}(d). Moreover, this PNL-induced
topologically protected phononic edge states for the armchair-edged
boundary of graphene has been further discussed in the supplementary
materials~\cite{SM}.

\begin{figure}
\includegraphics[width=0.47\textwidth]{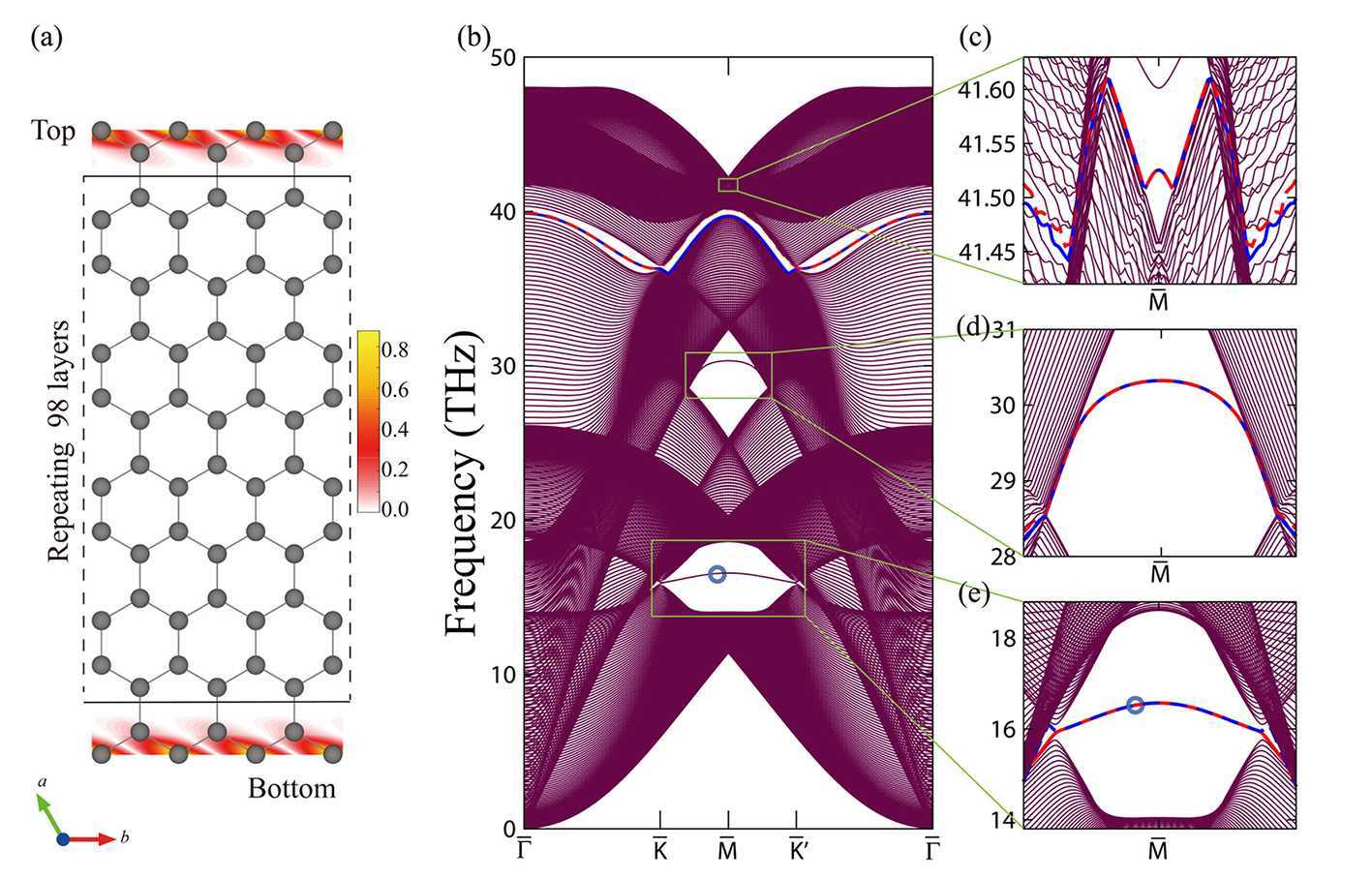}
\caption{The phonon dispersions of the zigzag-edged boundaries
(along the [100] direction) of graphene. Panel (a): The ribbon model
of graphene with the zigzag-edged boundary. Panel (b): The phonon
dispersion of the ribbon model derived by the tight-binding
calculations along the zigzag-edged high-symmetry paths. Panels
(c,d,e): the zoom-in local phonon dispersions connecting to the
projection sites of the PW3, PNL and PW1 phonons around the
$\overline{M}$ point. Note that we have plot the edge states in
Panel (a) on a selected point (as marked by a close circle in panel
(b)) of the PW1-induced topologically protected phononic edge
dispersion.} \label{fig:n4}
\end{figure}

The occurrence of single phononic Weyl-like nodes in the BZ is very
important for graphene, which provide an ideal platform to study
novel topological phononic properties, such as destructive
interference and quantum (anomalous/spin) Hall-like topological
effects. Firstly, it is clear that the chiral electrons in graphene
\cite{graphene2009} moving along a closed path have been
demonstrated to exhibit a phase change of the two components of the
wave function. This fact leads to a new phase, which contributes to
the interference processes. In similarity to electron in graphene,
the Weyl-like phonons at both $K$ and $K^\prime$ have the opposite
chirality for both PW1 and PW2. If phonon traverses such a closed
path without being scattered, protected by the PW-induced
non-trivial phononic edge states, from a PW at $K$ to the other PW
at $K^\prime$, the Berry phase definitely changes its sign of the
amplitude of one path with respect to the time-reversed
path\cite{graphene2009}. Therefore, these two paths possibly form
so-called destructive interference, as accompanied with a
suppression of backscattering. As shown in Fig.~\ref{fig:n4}(a) for
the robust phononic non-trivial edge states induced by the PW1 along
the \emph{K} to \emph{K}$^\prime$ path, the phonon destructive
interference would be reasonably expected. The similar behaviors
would also hold for both PW3 on the $\bar{\Gamma}$-$\bar{M}$
direction and PW4 along the $\bar{\Gamma}$-$\bar{K}$ direction.
Secondly, it is interesting to emphasize that, as shown in
Fig.~\ref{fig:n2}, the Berry curvatures are vanishingly zero
throughout the BZ, except for the positions of four pairs of PWs
(Fig.~\ref{fig:n2}). At these PWs, the non-zero Berry curvatures
(corresponding to the Berry phase of $\pi$ and $-\pi$) appear,
associated with PW1, PW2, PW3, and PW4 ( Fig.~\ref{fig:n2}). They
can be indeed viewed as magnetic field in momentum space,
accordingly leading to the possible occurrence of the phononic
quantum (anomalous/spin) Hall-like topological effects.

Unexpectedly, through first-principles calculations and modeling
analysis we have revealed the existence of intrinsic topological
Weyl-like and PNL phonons in pristine graphene, as accompanied with
the robust appearance of the topologically protected one-way
phononic edge states. Given the fact that many novel physical
properties are related with phonons, graphene would be an ideal case
to elucidate fundamental physical phenomena related with topological
phonons, possibly including heat conduction, electrical resistance,
and phonon wave-guides, as well as electron-phonon coupling effects
for superconductivity.

\bigskip
\noindent {\bf Acknowledgments} Work was supported by the National
Science Fund for Distinguished Young Scholars (No. 51725103) and by
the National Natural Science Foundation of China (Grant No.
51671193). All calculations have been performed on the
high-performance computational cluster in the Shenyang National
University Science and Technology Park.

\end{document}